# Constructing synthetic populations in the age of big data

M. A. Nicolaie[1], Koen Füssenich[2], Caroline Ameling[1], Hendriek C. Boshuizen[1]

1. RIVM (National Institute for Public Health and the Environment), Centre for Nutrition, Prevention and Health Services, P.O. Box 1, 3720, BA, Bilthoven, The Netherlands.

2. Capaciteit Orgaan, Domus Medica, Mercatorlaan 1200, 3525BL Utrecht, The Netherlands.

## Abstract

**Background**: To develop public health intervention models using microsimulations, extensive personal information about inhabitants is needed, such as socio-demographic, economic and health figures. Data confidentiality is an essential characteristic of such data, while the data should support realistic scenarios. Collection of such data is possible only in secured environments and not directly available for external micro-simulation models. The aim of this paper is to illustrate a method for construction of synthetic data by predicting individual features through models based on confidential data on health and socio-economic determinants of the entire Dutch population.

**Methods**: Administrative records and health registry data were linked to socio-economic characteristics and self-reported life-style factors. For the entire Dutch population (n = 16,778,708), all socio-demographic information except life-style factors was available. Life-style factors were available from the 2012 Dutch Health Monitor (n = 370,835). Regression model were used to sequentially predict individual features.

**Results**: The synthetic population resembles the original confidential population. Features predicted at the beginning of the sequential procedure are practically similar to those in the original population, while those predicted at the end of the procedure carry the limitation of previously modelled features.

**Conclusions**: By combining socio-demographic, economic, health and life-style related data at individual level on a large scale, our method provides us with a powerful tool to construct a synthetic population of good quality and with no confidentiality issues.

# Introduction

Given the rise in computing power, micro-simulation models of populations are increasingly used to support decision making in policy and fields such as epidemiology, demography, urban and environmental modelling (e.g. Devaux et al. (2020), Goryakin Y et al. 2020, Hendriksen et al. 2018)

An important part of individual based modelling is the construction of a starting population for the simulation (Beckman et al. (1996)). Next to being a starting point, they play a role in calibrating individual transition probabilities against population marginals. The distribution and associations of the relevant variables, such as baseline demographics, in this starting population should represent the real-world population as best as possible. Due to the increased digitalisation of society, more and more information is collected in digital form on all members of populations, and statistical offices can increasingly link this information to each other and to surveys. This information can be used to construct more accurate starting populations for modelling that do not depend on conditional independence assumptions that are implicit in some of the methods used before (e.g. Kooiker and Boshuizen, 2018).

An important limitation of the availability of large amounts of information on individuals in a population is that such data constitutes sensitive information, making privacy a primary issue.. For open use, the disclosure risk of such data needs to be low, while the statistical structure should be as realistic as possible. Although in many countries data are available for selected researchers, this is usually under strict conditions to minimize the risk of disclosure of information on each single individual. Running a model on real population data therefore generally will only be allowed within the secure environment (computer system) of the trusted party that manages these data. Running simulation models within these systems is generally not very practical, for instance due to lack of portability of software or lack of sufficient computer power. Furthermore, in the spirit of open science, it is preferable to make micro-simulation models, including their data, widely available for other scientists.

Construction of a synthetic population that can be used outside this secure environment can overcome these difficulties. Such a population reflects the structure of the available population data, but it is synthetic in the sense that individuals are constructed from this structure, and not represent real persons. Most papers on constructing synthetic population focus on taking multiple draws of individuals from a smaller survey population with detailed information in order to obtain a larger population (Alfons et al., 2010), Müller & Axhausen 2010, Barthelemy & Cornelis 2012), where the population constructed should be consistent with known demographics. Other approaches encompass synthetic reconstruction (Wilson and Pownall, 1976)), combinatorial optimization (Williamson (1998)) and model-based generation. The task in the first approach (upsampling smaller surveys) is to find inclusion probabilities for the survey individuals that deliver a population with equal marginal frequencies to those available from demographic sources. These methods largely overlap with methods for the construction of survey weights, that are used to make survey statistics representative for those of a target population. Others use simulation of past population history, where the microsimulation is run from the birth of each individual (a moment in the past) to deliver current prevalences of disease and risk factor histories (Kooier & Boshuizen, 2018).

This paper presents another methodology, in which regression models fitted on the original linked data are used to generate a synthetic population. We illustrate this for the case where we generate a Dutch population for chronic disease modelling. In chronic disease modelling one needs a starting population with known socio-demographic characteristics, exposures to risk factors for disease, and presence of chronic diseases.

The remainder of this paper is organized as follows: Section 2 provides the description of our method. Section 3 provides the evaluation of our method and Section 4 concludes the paper. Appendix A introduces the variables used and Appendix B illustrates the regression models used.

## Material and methods

The two main goals of the method were, on the one hand, the protection against disclosure risk for individuals that make up the original dataset and, on the other hand, to extract the available information from the original data for modelling purposes. We therefore constructed the synthetic population in a two-step procedure. The first step is carried out on microdata within the computer system of Statistics Netherlands (SN), under strict conditions regarding privacy. In this step we constructed predictive equations that do no longer contain information that can be traced back to an individual. In the second step these equations were used to construct a synthetic population outside the microdata environment.

The predictive equations in the first step were constructed as follows: starting from the original data set, we selected a small set of variables with no identity disclosure risks: age, gender, region and urbanicity further referred as "seed variables". We aimed to use a stratum sample size of minimally ten persons, that is, ten persons sharing the same combination of seed variables, following the Statistics Nederland guidelines on disclosure risk. Although there were a couple of strata with a lower number of individuals, the requirement was waved due to the insensitivity of the disclosed information (age, gender and location). The number of individuals in each stratum was then exported from the microdata environment.

Within the microdata environment, the seed variables were used as starting point to build regression models in a variable-by-variable approach, for a set of confidential variables (socio-demographic, lifestyle and presence of disease). The estimated model parameters as well as their covariance matrix were also exported from the microdata environment.

In the second step, we used the frequencies of the seed variable strata and the estimated predictive equations, both outside the microdata environment, for the generation of an artificial population.

In the following paragraphs, we will present each step in more detail. In Appendix A, we show in Table 1 the order in which the variables were included in the sequential modelling process, as well as the sources of these variables.

### Data sources

We targeted the population of the Netherlands at the 31st of December 2012, comprising 16 778 708 individuals (population size of the Netherlands on December 31$^{st}$ 2012). Our data sources were individual level, non-public, linkable microdata sets of Statistics Netherlands, made available under strict conditions regarding privacy issues. Virtually complete data were available on date of birth, gender, marital status, region (COROP code plus urbanity level), ethnicity, percentile group of household capital, source of income, percentile group of household income and household composition. Incomplete data on the highest level of education achieved were available for the non-institutionalized Dutch population aged 15 or older (see Appendix, Table 1).

Self-assessment of smoking, BMI and physical activity level was available for non-institutionalized individuals older than 18 years from survey data from the Dutch Public Health Monitor 2012 amounting to 387 195 participants (Van den Brink, C. et al, 2017). For this monitor, harmonized Health Surveys were conducted by the Municipal Health Services. The average participation rate was 47%. Several Municipal Health Services oversampled the elderly or those living in deprived areas, so that the sample as such is not representative for the Netherlands. However, as we fitted models conditioning on age and socio-economic factors, this is mitigated.

We addressed the issue of missing data in the targeted variables of the Dutch Health Monitor 2012 due to the person-level non-response by means of multiple imputation. We created five replications of complete risk factors data leading to the creation of five sets of regression coefficients. We pooled them using standard multiple imputation rules (Rubin, 1976).

Individual probabilities of having coronary heart disease (CHD), stroke, diabetes or chronic obstructive pulmonary disease (COPD) in 2012 were calculated from prediction models using demographic data and data on drug reimbursement. These prediction models were developed using as outcomes individual data on hospital admission and primary care use. In short, the construction of the prediction model involved using LASSO regression model for variable selection followed by a regular regression, as described in Füssenich, K. et al. (2021).

Incidence of lung- and pancreatic cancer were available from the Netherlands Comprehensive Cancer Organization (IKNL) cancer register, which records all individual cancer diagnoses for the Netherlands.

## Statistical prediction models

The set of seed variables selected from Statistics Netherlands were age, gender, COROP code (indicating regions) and urbanity. With a sequential approach we built a series of prediction models as follows: for each variable we fitted a prediction model only using as predictors the seed variables and the variables included in earlier prediction models. So, we started for the first outcome with a model that as predictors contains only the seed variables. For the second outcome, the predictors were the seed variables and the first outcome; for the third these were the seed variables, the first and the second outcomes, and so forth.

This approach has the potential to ensure accurate statistical properties (e.g., to preserve the moments of distribution and the associations between variables) for the selected confidential variables, if the fitted models capture the distributions and associations correctly. However, models are always limited and bias might be introduced when the models do not capture all relevant relations or distributions. The sequential nature of the procedure implies that the inaccuracies in the prediction from the first model will cause inaccuracies in all subsequent predictions. Therefore, we start with the variables we have for the entire Dutch population, and within this set with the variables that correlated highest with other variables.

We first modelled the main income source of the household on the seed population, as the main driver of the social heterogeneity in the population. Next, we continued with models for spendable household income and household capital, respectively, knowing that these can vary dramatically with the source of income. We extended the predictive equations by introducing successively type of household, household size and ethnic group. The reason for the introduction of the variable on education rather late in the sequential modelling approach is the fact that it was the only variable of the nation-wide registry data with a large amount of missingness (more than 40%, with missingness strongly depending on age).

After the nation-wide variables - with the exception of the presence of cancer - we added health and lifestyle information from the Dutch Health Monitor, such as BMI, smoking status and physical activity. After that we fitted models for the cancers. We chose this order because we considered lifestyle factors from the Dutch Health Monitor to be highly correlated with these types of cancer. In modelling these cancer types, we used data on the entire population, applying the missing indicator method (Rubin, 1987) for lifestyle factors, as only 2% of the population participated in the Dutch Health Monitor.

Given that diseases such as CHD, diabetes, stroke, CODP could be a consequence of exposure to risk factors, we decided to model them last in our sequence of models. These variables were expressed not as self-reported diagnosis registered in the Dutch Health Monitor, but as predicted probabilities from an earlier model (see Füssenich, K. et al., 2021), where the predictions were largely driven by the use of particular pharmaceutical drugs (a nation-wide available variable). Though these variables were available for the whole Dutch population, they were used in models fitted on the Dutch Health Monitor data only. These models completed the chain of predictive equations.

In appendix A, we describe in detail all these fitted models. Following the observation that many variables depended on age in a non-linear fashion, we used spline functions of age. As several variables had interaction with gender, we fitted all models stratified by gender and, when necessary, by other variables. As a general feature of the modelling approach, we restricted ourselves to models with main effects only.

## Construction of initial population

From the seed population exported from the confidential environment we constructed the synthetic population of size 16 768 952 (after listwise removal of participants with missing data) by randomly drawing from the predictive equations as follows. We started with the first predictive equation on the seed population and generate the first synthetic variable. Notably, these generated values do not contain real data for the targeted variable. Next, we sequentially simulated from the subsequent predictive equations. We limited the construction to the ages 0-105 years, as these are generally the ages to include in simulation models.

## Methods of evaluating the results

As a method to investigate the utility of our approach, we compared the statistical attributes of the synthetic population with those of the confidential original population using univariate and multivariate statistics. For the univariate results, we used the generated frequencies for categorical variables and the first four moments of risk factors distribution for continuous variables. For the multivariate results, we used joint distributions stratified by age class, gender, smoking status and educational level. For this stratification, we recoded age as an eight-level categorical variable (from 20 to 80 years in 10-years age classes, with two extra categories for younger than 20 and older than 80).

## Results

For univariate results, in Table 3a, we contrast frequencies for all categorical variables within the confidential original population and the synthetic population. In Table 3b, we contrast the first four moments of continuous variables from the two populations.

In general, the frequencies of the generated data are close to those of the original data, although for the number of persons in the household and type of household the numbers were slightly different between synthetic and original. For lung cancer the prevalence was slightly lower in the synthetic population.

To get an impression on lifestyle variables, which were not available for the entire population, in Tables 4a and 4b we compare the generated data for people 19 years and older with those in the Dutch Health Monitor population. However, this comparison should be made with caution as the latter is not a representative sample, as illustrated the differences in age in table 4b. This also drives much of the differences seen in disease probabilities in Table 4a. The proportion of smokers in the synthetic population (22.4%) of age 19 and older is lower than the 22.8% reported from the Dutch Health Monitor for the same period (see PHPinfo website).

Comparing standard deviation, skewness and kurtosis, we see that the standard deviation of BMI is similar in the synthetic population and the original data. However, skewness and kurtosis differ. As the model used assumes normally distributed residuals, skewness and kurtosis of the generated data are close to zero, while they are larger for the original data.

To get a better idea on how the predictions worked within strata, we also made figures by age, gender, smoking and educational group. As smoking is available only from the Dutch Health Monitor data, again this compares the prevalence in the constructed population with that in the population participating in the Dutch Health Monitor. However, now we look at differences within strata of age, gender and education level eliminating differences due to those variables. Figures 1 and 2 indicate that in the synthetic population the mean BMI seems to be overestimated in lower age groups and in women with lower or higher tertiary education. Figures 3 and 4 indicate that mean physical activity is generally correctly estimated in the synthetic population compared to the Dutch Health Monitor population with the exception of the 80+ population, where it was estimated to be higher.

Figures 5 and 6 show that lung cancer prevalence is slightly lower in the lower educated in the synthetic population than in the confidential data.

Figures 7 and 8 show that smoking prevalence seems to be reconstructed reasonably well, but the separation of non-smokers into former smokers and never smokers in the younger age groups differs considerably from the data in the Dutch Health Monitor. In the smoking model, demographic variables such as ethnicity and region have relatively large coefficients, and might be differently distributed in the general population compared to the Dutch Health Monitor population. For instance, those with a non-western background comprised 4.5% of the Dutch Health Monitor population, but 11.8% of the Dutch population..

Table 3a. Frequencies (in percent) of categorical variables in the confidential original population and in the synthetic population.

| Variables | Confidential original total population | Synthetic population |
|---|---|---|
| | Frequencies | Frequencies |
| **Lung cancer** | | |
| no | 99.86 | 99.88 |
| yes | 0.14 | 0.12 |
| **Pancreas cancer** | | |
| no | 99.99 | 99.99 |
| yes | 0.01 | 0.01 |
| **Source of income** | | |
| Employee | 47.0 | 47.0 |
| Civil servant | 7.4 | 7.4 |
| Salary as company director | 2.4 | 2.4 |
| Other income from labour | 0.3 | 0.3 |
| Income as company owner | 14.7 | 14.7 |
| Income from property | 0.4 | 0.4 |
| Unemployment benefits | 1.0 | 1.0 |
| Disability pension | 2.9 | 2.9 |
| Retirement pension | 17.8 | 17.8 |
| Social assistance benefits | 3.2 | 3.2 |
| Other social security | 1.0 | 1.0 |
| Study grant | 0.8 | 0.8 |
| Other | 0.1 | 0.1 |
| no income | 1.0 | 1.0 |
| **Household size (number of persons)** | | |
| 1 | 17.5 | 17.2 |
| 2 | 29.7 | 30.2 |
| 3 | 16.4 | 16.7 |
| 4 | 23.1 | 22.4 |
| 5 | 9.4 | 9.4 |
| 6 and more | 3.9 | 4.1 |
| **Migration background** | | |
| Dutch | 78.9 | 78.7 |
| Moroccan | 2.2 | 2.2 |
| Turkish | 2.4 | 2.4 |
| Surinam | 2.1 | 2.1 |
| Netherlands Antilles and Aruba | 0.9 | 0.9 |
| Other non-Western | 4.2 | 4.2 |
| Other Western | 9.4 | 9.5 |
| **Type of household** | | |
| institutional | 1.4 | 1.6 |
| non-institutional | 98.6 | 98.4 |

Table 3b. Summary statistics of continuous variables in the confidential original population and in the synthetic population.

| Variables | Confidential original population | | | | Synthetic population | | | |
|---|---|---|---|---|---|---|---|---|
| | Mean | Standard deviation | Skewness | Kurtosis | Mean | Standard deviation | Skewness | Kurtosis |
| Property | 50.49 | 30.18 | -0.11 | -1.25 | 49.76 | 29.67 | -0.04 | -1.24 |
| Income | 59.43 | 27.39 | -0.44 | -0.85 | 58.41 | 27.85 | -0.43 | -0.93 |
| Age[1] | 40.29 | 22.96 | 0.075 | -0.93 | 40.29 | 22.96 | 0.075 | -0.93 |

1) The original population contained 9756 subjects not included in the synthetic population due to missing data.

Table 4a. Summary statistics of categorical variables in the confidential original Dutch Health Monitor population and in the synthetic population for individuals aged 18+.

| Variables | Dutch Health Monitor (age > 18) | Synthetic population (age > 18) |
|---|---|---|
| | Frequencies | Frequencies |
| **Smoking** | | |
| Never smoker | 41.6 | 26.9 |
| Past smoker | 41.4 | 50.7 |
| Light smoker | 13.3 | 16.3 |
| Heavy smoker | 3.7 | 6.1 |
| **Physical activity (complies with norms)** | | |
| no | 34.1 | 37.1 |
| yes | 65.9 | 62.9 |
| **Education (completed) [SOI level]** | | |
| Primary or less [1,2] | 12.8 | 8.4 |
| Lower secondary [3-6] | 18.8 | 13.5 |
| Higher secondary [7-10] | 41.8 | 53.3 |
| Lower Tertiary [11-13] | 18.2 | 16.3 |
| Higher tertiary [14+] | 8.4 | 8.3 |
| **Diabetes present** | 11.5 | 5.8 |
| **COPD present** | 6.6 | 7.9 |
| **CHD present** | 10.5 | 2.7 |
| **Stroke present** | 5.4 | 4.4 |

Table 4b. Summary statistics of continuous variables in the confidential Public Health Monitor population and in the synthetic population for individuals age 19 and above. Note that the composition of the survey population differs from that of the total population

| Variables | Dutch Health Monitor (age > 18) | | | | Synthetic population (age > 18) | | | |
| --- | --- | --- | --- | --- | --- | --- | --- | --- |
| | Mean | Standard deviation | Skewness | Kurtosis | Mean | Standard deviation | Skewness | Kurtosis |
| Age | 57.06 | 17.88 | -0.34 | -0.79 | 48.98 | 17.91 | 0.22 | -0.76 |
| BMI | 25.73 | 4.09 | 0.87 | 1.43 | 25.71 | 4.06 | -0.04 | 0.13 |
| Property | 60.20 | 28.39 | -0.56 | -0.79 | 51.25 | 29.67 | -0.10 | -1.23 |
| Income | 58.10 | 25.89 | -0.26 | -0.94 | 56.57 | 28.00 | -0.35 | -1.00 |

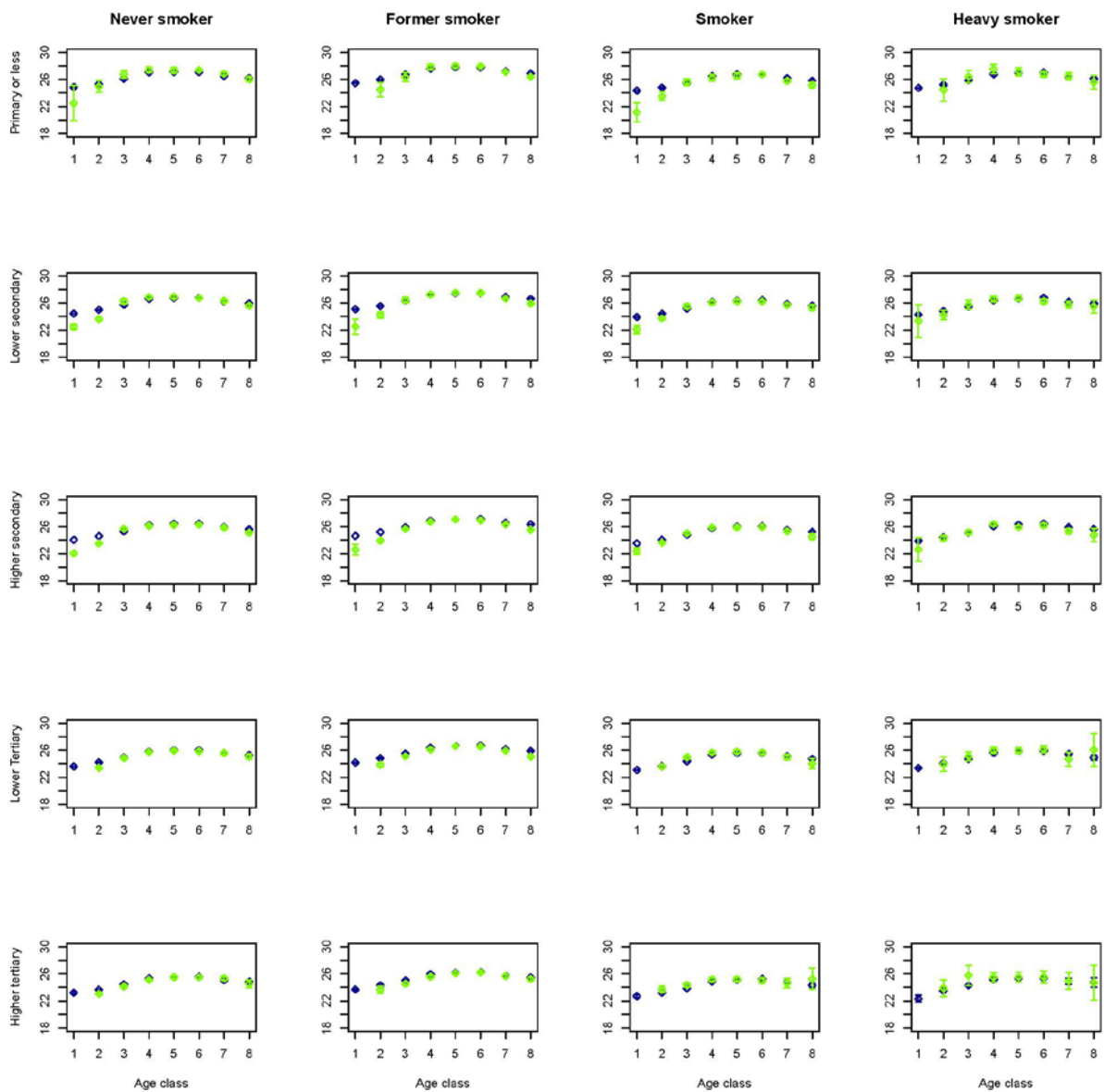

*Figure 1 Estimated BMI mean and their 95% CI for men in each 10 years age class (1=<20 ,2=20-29,  3=30-39, 4=40-49, 5=50-59, 6=60-69, 7=70-79, 8=80+) by smoking status and educational level. Blue is the synthetic population, green the original data from the Dutch population*

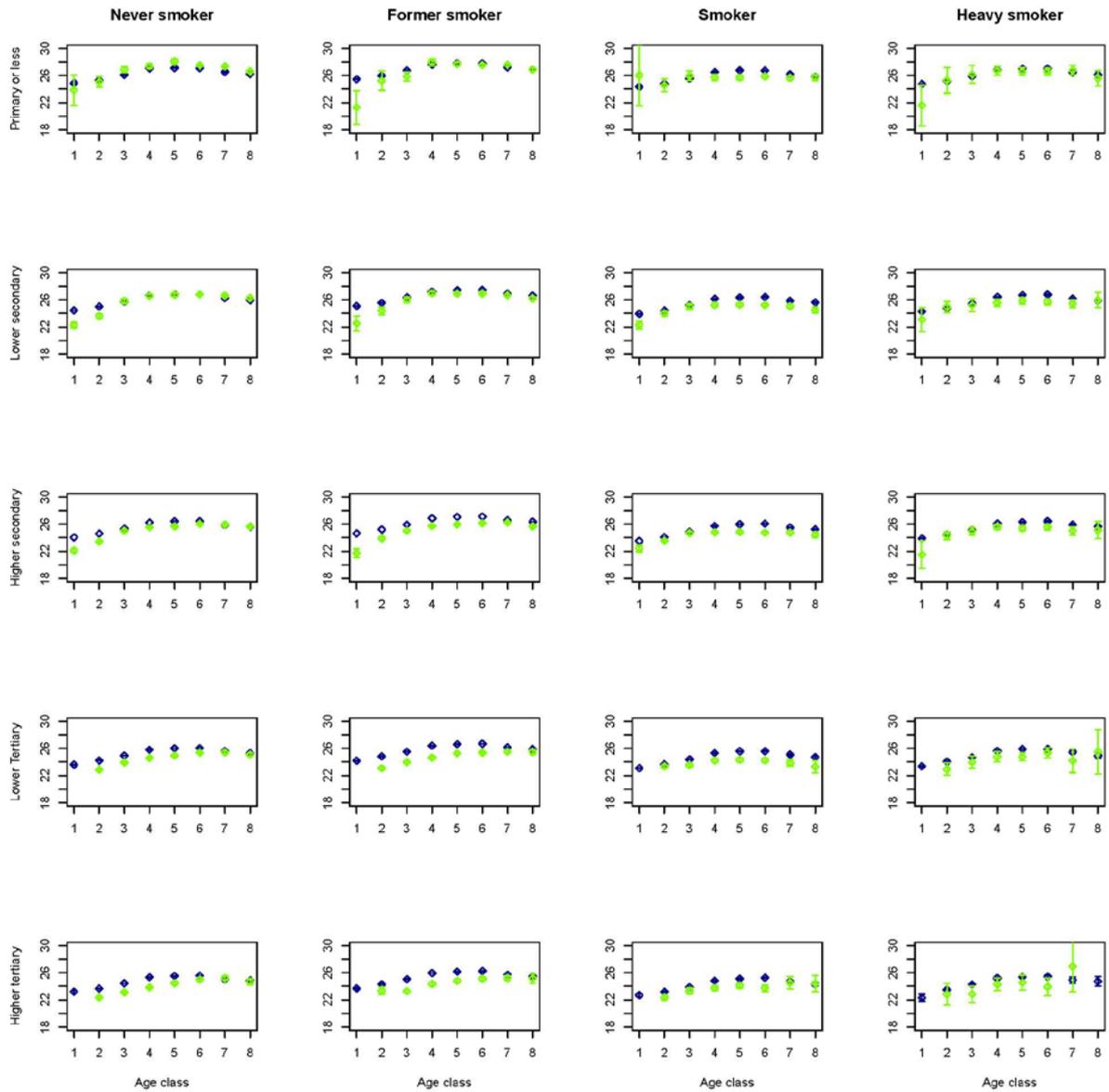

*Figure 2 Estimated BMI mean and their 95% CI for women in each 10 years age class (1=<20 ,2=20-29, 3=30-39, 4=40-49, 5=50-59, 6=60-69, 7=70-79, 8=80+) by smoking status and educational level. Blue is the synthetic population, green the original data from the Dutch population*

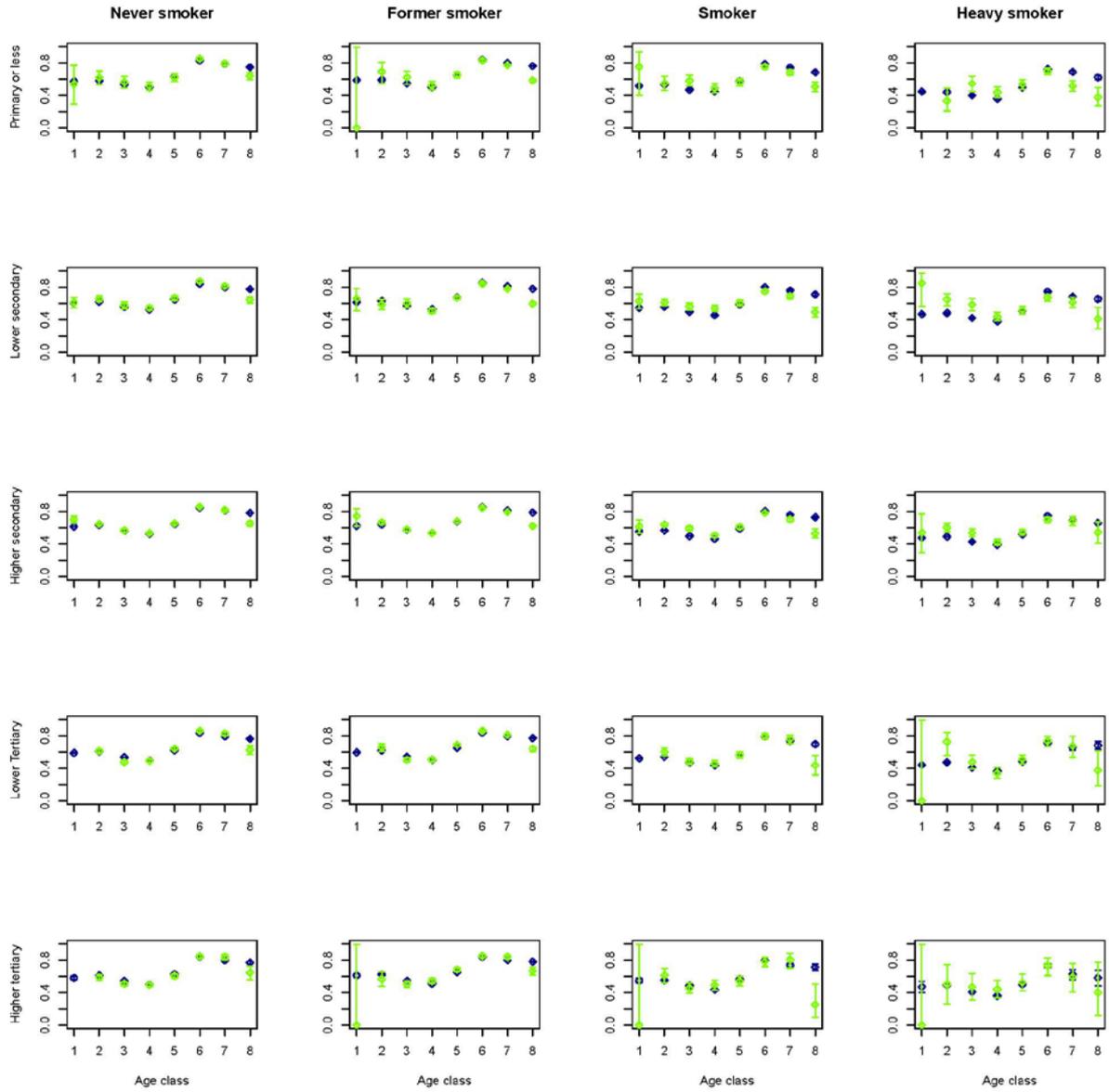

*Figure 3: Prevalence of being sufficiently physical active and 95% CI for men in each 10 years age class (1=<20 ,2=20-29, 3=30-39, 4=40-49, 5=50-59, 6=60-69, 7=70-79, 8=80+) by smoking status and educational level. Blue is the synthetic population, green the original data from the Dutch Public Health Monitor 2012*

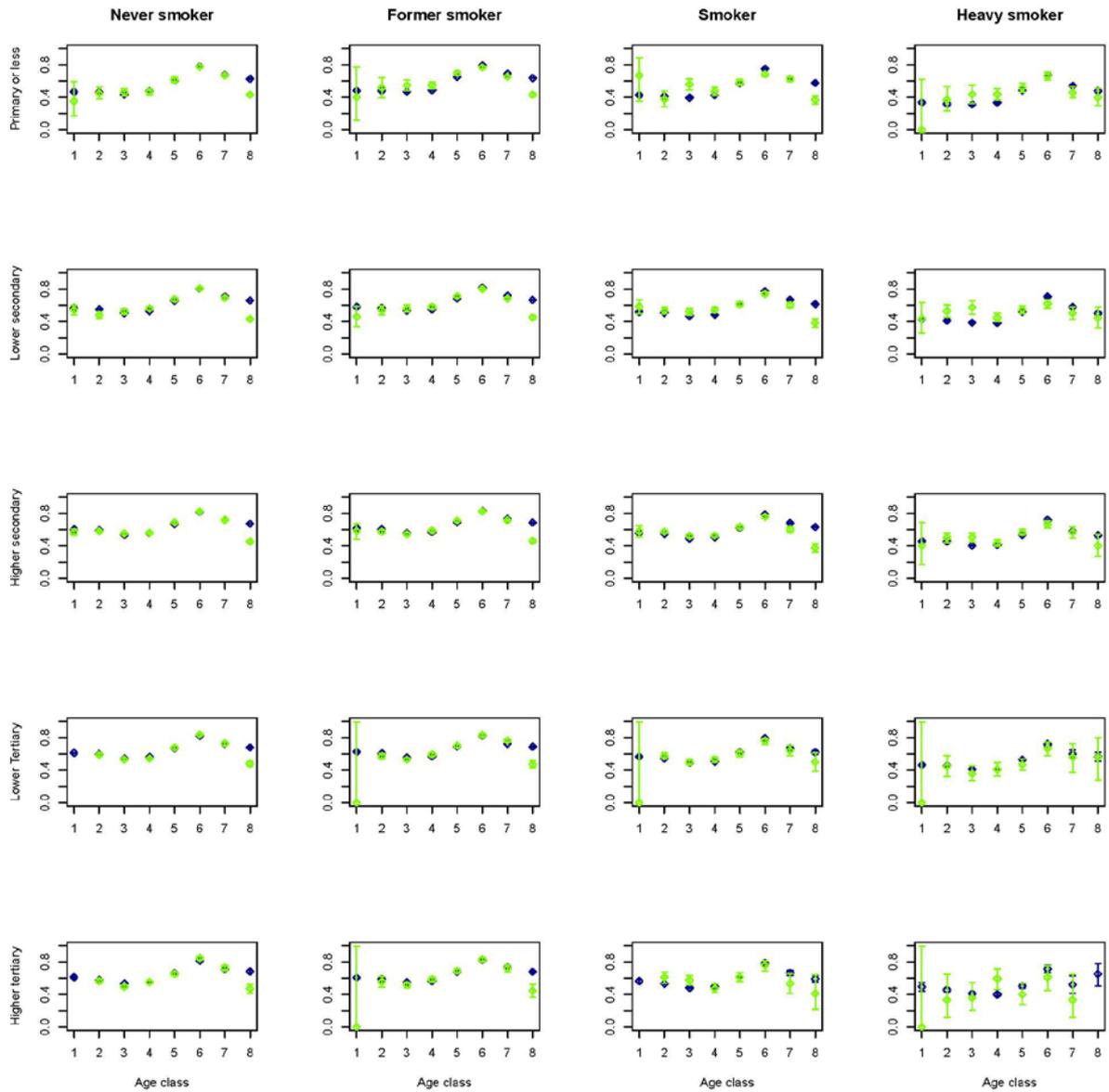

*Figure 4 Estimated mean physical activity and 95% CI for women in each 10 years age class (1=<20, 2=20-29, 3=30-39, 4=40-49, 5=50-59, 6=60-69, 7=70-79, 8=80+) by smoking status and educational level. Blue is the synthetic population, green the original data from the Dutch Public Health Monitor 2012*

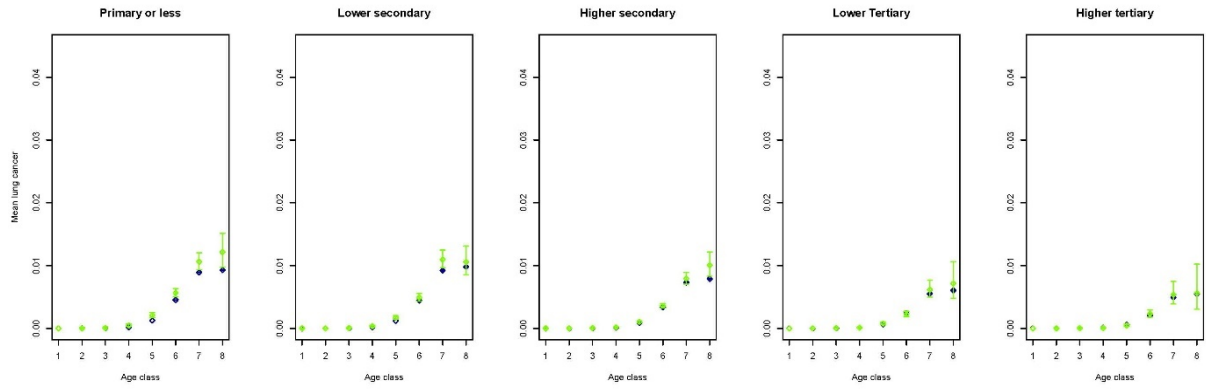

*Figure 5 Estimated mean of lung cancer and their 95% CI for men in stratum education level across 8 age classes (1=<20 ,2=20-29, 3=30-39, 4=40-49, 5=50-59, 6=60-69, 7=70-79, 8=80+). Blue stands for the synthetic population, green stands for the original data from the cancer registration linked to the Dutch Public Health Monitor 2012*

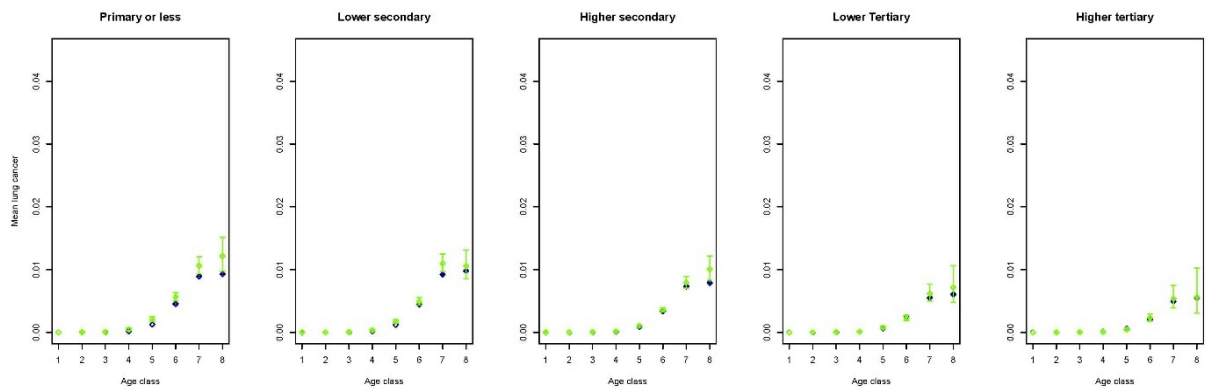

*Figure 6 Estimated mean of lung cancer and their 95% CI for women in stratum education level across 8 age classes (1=<20 ,2=20-29, 3=30-39, 4=40-49, 5=50-59, 6=60-69, 7=70-79, 8=80+). Blue stands for the synthetic population, green stands for the original data from the cancer registration linked to the Dutch Public Health Monitor 20*

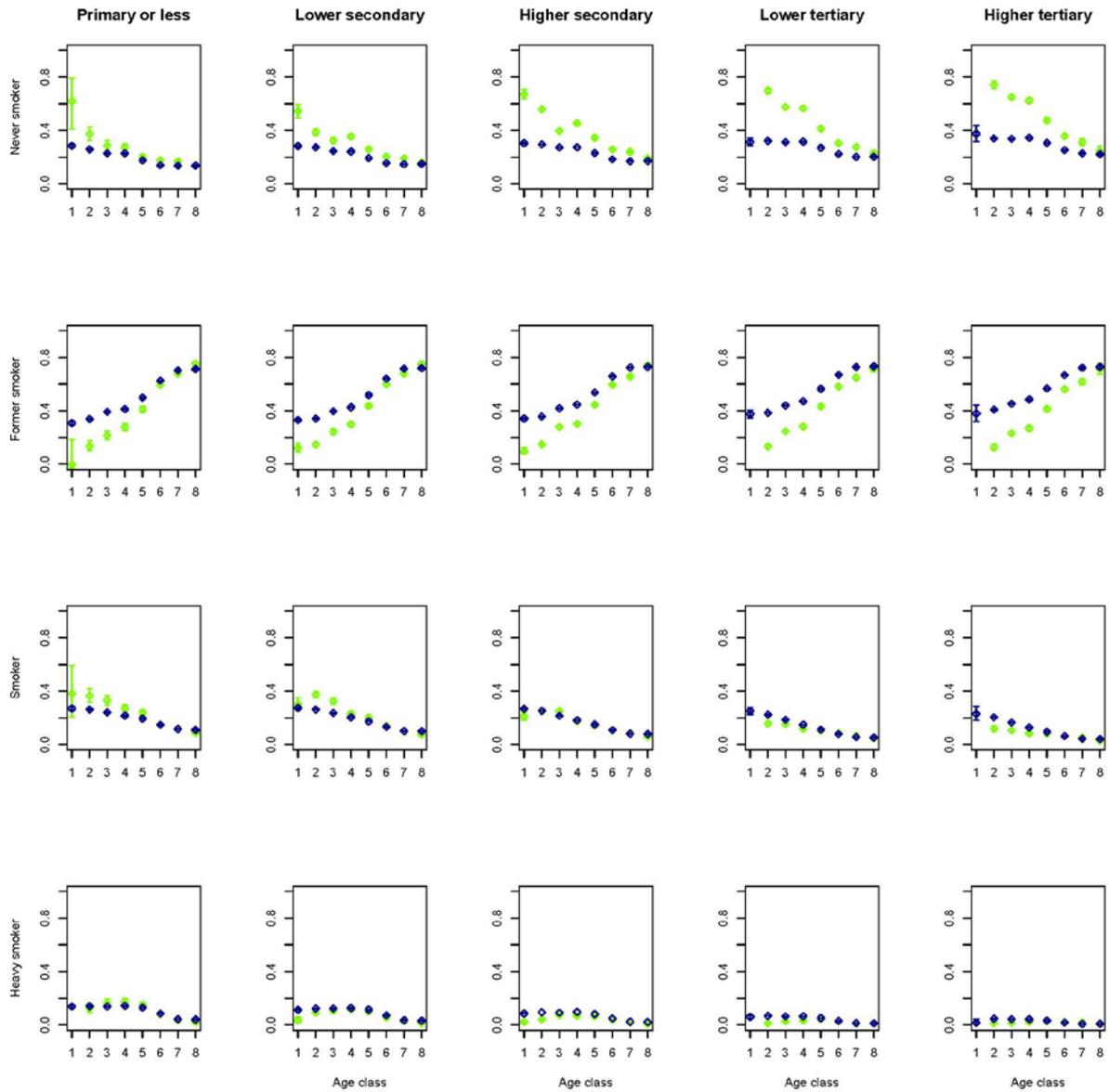

Figure 7 Percentage of smoker categories across 8 age classes (1=<20 ,2=20-29,  3=30-39, 4=40-49, 5=50-59, 6=60-69, 7=70-79, 8=80+) and education level for men based on original data from of the Dutch Public Health Monitor 2012.

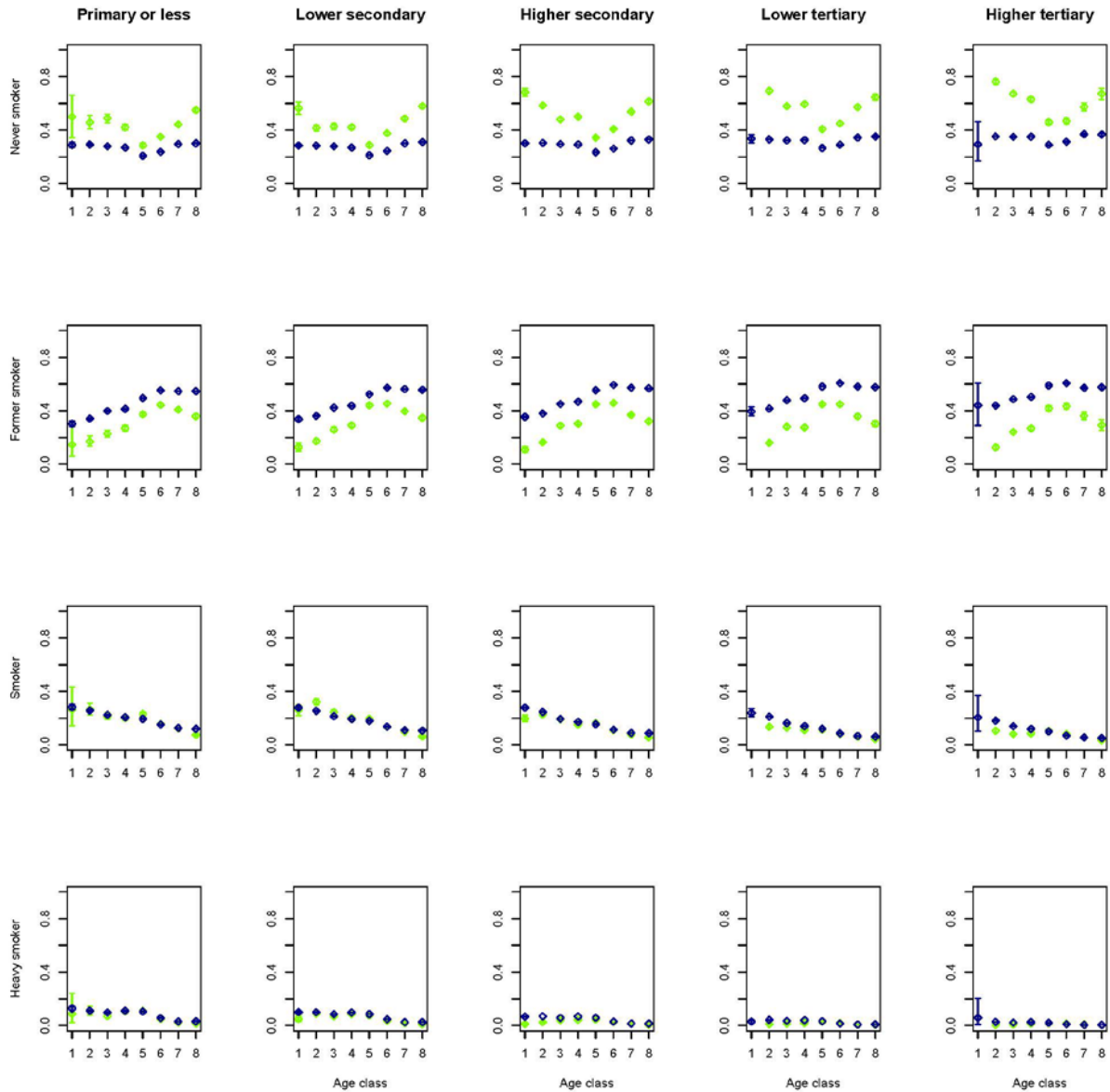

*Figure 8 Percentage of smoker categories across 8 age classes (1=<20 ,2=20-29,  3=30-39, 4=40-49, 5=50-59, 6=60-69, 7=70-79, 8=80+) and education level for women based on original data from of the Dutch Public Health Monitor 2012*

## 4. Discussion

The objective of this paper is developing a methodology to generate a synthetic population for application in chronic disease modelling, based on linked data at the level of individuals, household's demographics and lifestyle characteristics in the presence of disclosure risk. A synthetic population amounting to the size of the total Dutch population of 2012 was created with realistic characteristics. Our method relies on the construction, in a sequential manner, of regression models for the distributions of individual attributes conditional on a set of determinants. It then uses these regression models to simulate a population by drawing from them. Each synthetic individual has a probabilistically assigned value for all but the initial determinants age, gender, region and urbanicity, which are taken from the original data.

In our application we had access to rich individual level data, for instance household capital and income, based on Tax Authority data.

Part of our data were virtually complete (household information, demographic information, and income and property information). This yields, on the one hand, the benefit of dealing with minimum uncertainties in the modelling of these variables. On the other hand, model selection was time consuming, as the choice of potential models is large and running times on 16 million records are substantial, specifically when using complex models. As our approach needs parametric models, more automated prediction methods like random forest or other ensemble machine learning approaches could not be used. We settled for relatively simple models, assuming for instance normal residuals and only including a minimum of interaction terms.

This means that the generated data reflected these assumptions, and the generate BMI data are close to being normally distributed, while the original data were not. Whether this is a problem will depend on how the synthetic population is further used. For uses as an initial population in micro-simulation of disease we do not deem this problematic. If the skewness of the original data needs to be reflected, the method can be easily adapted by using a model that better reflects the skewness of the data.

An analysis of the goodness-of-fit of the synthetic to the true population using summary statistics and qq-plots showed that it was possible to achieve a high degree of accuracy for the set of first 7 variables (from age to ethnic group), given the availability of the true, complete population. For the next variables, based on the Dutch Health Monitor data, validation of the results was difficult, as the population of the Dutch Health Monitor has a different composition as the total Dutch populations. For instance, in the synthetic population the number of former smokers was considerably higher and the number of never smokers lower than in the Dutch Health Monitor population, despite the latter being younger. This means that the Dutch Health Monitor population has a composition that deviates importantly from the general population. Therefore, the stratified comparisons we conducted are more informative. These comparisons show better similarities, but several disparities were noted, that are probably due to insufficient use of interaction terms in the modelling. Further validation could be done using information from other sources (Van der Steen, 2016), or by using survey weights on the original data. This is especially valuable where comparison against the original raw data is not very meaningful, as is the case for the data derived from the Dutch Health Monitor.

For prevalence of lung and pancreas cancer we had population wide data, but models were partly dependent on variables (like smoking) that came from a Dutch Health Monitor data, that covered only 2% of the population. Lung cancer prevalence was slightly lower as it should be. This might be due to the way the model was constructed, which is vulnerable to residual confounding: the missing indicator method used implies that effects estimated for age or gender are only adjusted for smoking in the 2% individuals that participated in the Dutch Health Monitor survey. So, most of these effects will not be adjusted, while the construction method assumes they are adjusted. Using only the Health Monitor population for fitting a lung cancer model would produce models that are less biased. However, given the low prevalence of lung cancer in the population, such models would be very imprecise. Given this discrepancy, for use in microsimulation modelling we recommend doing a post-calibration step, by increasing the probability of lung cancer during generation of the synthetic population with a factor that makes the prevalence equal to the population marginal.

An important limitation of our modelling approach might be grounded in the fact that we did not include interactions in our models, apart from stratification. As a result, potential interaction become part of the error term, magnifying the unexplained variance. We have observed this in the modelling of household variables, where the relation between personal capital and income was not well described by a simple additive relationship, even when using splines. A more adequate method could be non-parametric models like random forest, but exporting such models comes with disclosure risk and, therefore, are unfeasible in this context. However, as we experienced that manual model selection is time consuming, finding exportable models that automatically include or select important interactions would be a nice avenue to explore in further research.

Due to the large amount of information, challenges were faced in each model as the fit could always be improved. For instance, in the extremes (lowest, highest) of spendable income ranges, the relationship with personal capita is generally curvilinear, while in the middle-income range the relationship is linear. The explanation of this phenomenon at the lower end is partly that in some sectors (like agriculture) high investments (e.g. in land) are needed while income can be low. This relation could be the subject of further methodological and substantive investigations.

We now constructed the synthetic population using the point estimates of the parameters. However, we also exported the covariance matrices of the parameter estimates, so it would be easy to use this to randomly draw parameter sets multiple times and construct sets of synthetic populations, each based on a random draw. These can be used to estimate the influence of the statistical uncertainty in the models on the outcomes of the microsimulation.

In our approach, all data apart from the seed variables are generated from fitted models. This might be too cautious with regards to disclosure risks. It would be interesting to see whether it is possible to development a method which mixes real individuals with synthetic individuals and partially synthetic individuals in a manner that has no risk of disclosure.

In this paper we described the methods we used to construct a synthetic population meant for microsimulation, based on population wide data liked to data of a large health survey. Although the constructed population does not reflect the original data exactly, we believe the reconstruction is close enough to reality to be suitable for use in microsimulation. The method can be further improved by using models that include more interactions and transformations of dependent variables with non-normal residuals. It is surely superior to methods constructing initial populations that assume independence between variables. When marginals deviate too much from the available population marginals – mostly for variables constructed late in the construction procedure – calibration can be added to adjust marginals, while maintaining the mutual relations between variables assumed by the models.

## Appendix A: Description of variables

Table 1. The list of variables and their order in the sequential construction of synthetic population.

| Variable | Definition | main source | Method of collection | Missing (%) |
| --- | --- | --- | --- | --- |
| COROP region | a regional area within the Netherlands | Statistics Netherlands | Definition by SN | 0 |
| Urbanity | ambient/area address density | Statistics Netherlands | Definition by SN | 0 |
| Gender | Indicator | Civic Register | Municipality | 0 |
| Age | Age at 31st December 2012 | Civic Register | Municipality | 0 |
| Source of income | Main source of income in the household | Tax Authorities, DUO | Deducted by SN? | 0 |
| Spendable income | Percentile of spendable income of the Dutch household | Tax Authorities, DUO | Calculated by SN | 0 |
| Personal capital | Percentile of personal capital of the Dutch household | Tax Authorities | Calculated by SN | 0 |
| Type of household | Type of household at the beginning of the year | Civic Register | Derived by SN | 0 |
| Number of persons in the household | Household size | Civic Register | Derived by SN | 0 |
| Ethnic group | Categories based on country of origin | Civic Register | Municipality | 0 |
| Education | SOI classification | Statistics Netherlands | Defined by SN | 42.3 |
| Smoking | Smoking status | RIVM/ Statistics Netherlands | Self-assessment | 0 |
| BMI | Body mass index | RIVM/ Statistics Netherlands | Self-assessment | |
| Physical Activity | Adherence to PA norm | RIVM/Statistics Netherlands | Self-assessment | 0 |
| CHD | Probability | combined health care records* | Calculated by RIVM | 0 |
| Stroke | Probability | combined health care records* | Calculated by RIVM | 0 |
| Diabetes | Probability | combined health care records* | Calculated by RIVM | 0 |
| COPD | Probability | combined health care records* | Calculated by RIVM | 0 |
| pancreatic cancer | Diagnosed | Netherlands Comprehensive Cancer Organization (IKNL) | Reported by doctor | 0 |
| lung cancer | Diagnosed | Netherlands Comprehensive | Reported by doctor | 0 |

|  |  | Cancer Organization (IKNL) |  |  |

# Appendix B: Description of models

The starting population to support the models is the complete 2012 Dutch population whose retained characteristics are the seed variables: age (in years), gender, and place of residence at the level of COROP code (40-level variable) and urbanity rank. Following exploratory analysis, several observations which characterize all models are notable. All models are stratified per gender. Given that the effects of age on the outcomes vary largely in shape and magnitude, we employ natural cubic splines of age to model these effects with knots at 0, 10, 17, 20, 25, 30, 50, 55, 60, 66, 70, 80 90 and 100 for each stratum in turn. As dependent variables, we transform the percentile of spendable income and the percentile of personal capital through a linear transformation into z-scores. As independent variables, we transform the z-scores of the percentile of spendable income and the percentile of personal capital in natural cubic splines with knots at -2,-1, 0, 1 and 2 respectively, for each stratum in turn. The performance of model fit is assessed by 10-fold cross-validation.

Model 1 employs main income source of the household (categorical, 14 levels) as dependent variable and the seed variables as independent variables. We stratified at the intersection of gender and COROP region. We fit 80 multinomial regression models with main effects of urbanity and cubic splines.

Model 2 employs the percentile of spendable income as dependent variable and the seed variables and the main income source of the household as independent variables. We fit a linear regression models with main effects of COROP region, urbanity, cubic splines and main income source of the household.

Model 3 employs the percentile of personal capital as dependent variable and the seed variables, the main income source of the household and the percentile of spendable income as independent variables. We fit a linear regression models with main effects of the independent variables and natural cubic splines.

Model 4 employs the type of household (categorical, 2 levels) as dependent variable and the seed variables, the main income source of the household, the percentile of spendable income, the percentile of personal capital and the number of persons in the household as independent variables. We fit a multinomial regression models with main effects of the independent variables and natural cubic splines.

Model 5 employs the number of persons in the household (categorical, 6 levels) as dependent variable and the seed variables, the main income source of the household, the percentile of spendable income and the percentile of personal capital as independent variables. We stratify for gender and type of household. We fit a multinomial regression models with main effects of the independent variables and natural cubic splines.

Model 6 employs the ethnic group as dependent variable (8-level category) and the seed variables, the main income source of the household, the percentile of spendable income, the percentile of personal capital, the number of persons in the household and the type of household as independent variables. We stratify at the intersection of gender and type of household. We fit 8 multinomial regression models with main effects of the independent variables and natural cubic splines.

Model 7 employs education (categorical, 5 levels) as dependent variable and the seed variables, the main income source of the household, the percentile of spendable income, the percentile of personal capital, the number of persons in the household, the type of household and the ethnic group as independent variables. We fit a multinomial regression models with main effects of the independent variables and natural cubic splines.

Model 8 employs smoking status (categorical, 4 levels: "never smoker", "former smoker", "current smoker" and "heavy smoker") as dependent variable and the seed variables, the main income source of the household, the percentile of spendable income, the percentile of personal capital, the number of persons in the household, the type of household, the ethnic group and education as independent variables. We fit a multinomial regression models with main effects of the independent variables and natural cubic splines.

Model 9 employs BMI (continuous) as dependent variable and the seed variables, the main income source of the household, the percentile of spendable income, the percentile of personal capital, the number of persons in the household, the type of household, the ethnic group, education and smoking as independent variables. BMI was measured for individuals older than 18 years. We fit a linear regression models with main effects of the independent variables and natural cubic splines.

Model 10 employs physical activity (categorical, 2 levels) as dependent variable and the seed variables, the main income source of the household, the percentile of spendable income, the percentile of personal capital, the number of persons in the household, the type of household, the ethnic group, education, smoking and BMI as independent variables. We fit a multinomial regression models with main effects of the independent variables and natural cubic splines.

Model 11 employs pancreas cancer as dependent variable and the seed variables, the main income source of the household, the percentile of spendable income, the percentile of personal capital, the number of persons in the household, the type of household, the ethnic group, education, smoking, BMI and physical activity as independent variables. We fit logistic regression models with main effects of the independent variables and natural cubic splines for the z-scores of percentiles of personal income and personal capital. The missing data indicator approach has been use to deal with the missing values of smoking, BMI and physical activity for those residents not participating to the Health Survey study Missing values of education, smoking, BMI and physical activity within the Health Survey study have a priori been imputed five times; each imputed data sample has been merged with the administrative data of the entire Dutch population.

Model 12 employs lung cancer as dependent variable and the seed variables, the main income source of the household, the percentile of spendable income, the percentile of personal capital, the number of persons in the household, the type of household, the ethnic group, education, smoking, BMI, physical activity, and pancreas cancer as independent variables. We fit logistic regression models with main effects of the independent variables and natural cubic splines for the z-scores of percentiles of personal income and personal capital. A similar missing data indicator approach has been use to deal with missing values of education, smoking, BMI and physical activity as for the case of pancreas cancer.

Model 13 employs CHD as dependent variable and the seed variables, the main income source of the household, the percentile of spendable income, the percentile of personal capital, the number of persons in the household, the type of household, the ethnic group, education, smoking, BMI, physical activity, lung cancer and pancreas cancer as independent variables. We fit linear regression models on the logistic transformation of CHD with main effects of the independent variables and age as factor.

Model 14 employs stroke as dependent variable and the seed variables, the main income source of the household, the percentile of spendable income, the percentile of personal capital, the number of persons in the household, the type of household, the ethnic group, education, smoking, BMI, physical activity, lung cancer, pancreas cancer and CHD as independent variables. We fit linear regression models on the logistic transformation of stroke with main effects of the independent variables and age as factor.

Model 15 employs diabetes as dependent variable and the seed variables, the main income source of the household, the percentile of spendable income, the percentile of personal capital, the number of persons in the household, the type of household, the ethnic group, education, smoking, BMI, physical activity, lung cancer, pancreas cancer, CHD and stroke as independent variables. We linear regression models on the logistic transformation of diabetes with main effects of the independent variables and age as factor.

Model 16 employs COPD as dependent variable and the seed variables, the main income source of the household, the percentile of spendable income, the percentile of personal capital, the number of persons in the household, the type of household, the ethnic group, education, smoking, BMI, physical activity, lung cancer, pancreas cancer, CHD, stroke and diabetes as independent variables. We fit linear regression models on the logistic transformation of COPD with main effects of the independent variables and age as factor.